\documentclass[preprint,amsmath,aps]{revtex4}
\usepackage{graphicx}
\usepackage{dcolumn}
\usepackage{bm}
\usepackage{enumerate} 
\usepackage{hhline}
\begin{document}

\title{Exploring evidence of interaction between dark energy and dark matter}

\author{Daniela Grand\'on}
 \email{daniela.grandon@alumnos.uv.cl}
\author{V\'ictor H. C\'ardenas}
\email{victor.cardenas@uv.cl}
\affiliation{
Instituto de F\'isica y Astronom\'ia, Facultad de Ciencias, Universidad de Valpara\'iso, Av. Gran Breta\~na 1111, Valpara\'iso, Chile }

\date{\today}

\begin{abstract}
In this work we use the latest observations on SNIa, $H(z)$, BAO, $f_{gas}$ in clusters and CMBR, to constrain three models showing an explicit interaction between dark matter and  dark energy. In particular, we use the BOSS BAO measurements at $z \simeq 0.32$, $0.57$ and $2.34$, using the full 2-dimensional constraints on the angular and line of sight BAO scale. We find that using all five observational probes together, two of the interaction models show positive evidence at more than 3 $ \sigma $. Although significant, further study is needed to establish this statement firmly.
\end{abstract}

                        
\maketitle

\section{Introduction}

One of the most important problems of theoretical physics is to
explain the fact that the universe is in a phase of accelerated
expansion. Since 1998 \cite{Riess:1998cb}, \cite{Perlmutter:1998np}, the physical
origin of cosmic acceleration remains a deep mystery. According to
general relativity (GR), if the universe is filled with ordinary
matter or radiation, the two known constituents of the universe,
gravity should slow the expansion. Since the
expansion is speeding up, we are faced with two possibilities,
either of which would have profound implications for our
understanding of the cosmos and of the laws of physics. The first is that 75\%
of the energy density of the universe exists in a new form with
large negative pressure, called dark energy (DE) \cite{Peebles:1987ek}, \cite{Ratra:1987rm}, \cite{Wetterich:1987fm}. The other possibility
is that GR breaks down on cosmological scales and
must be replaced with a more complete theory of gravity \cite{Joyce:2016vqv}. In this
paper we consider the first option. The cosmological constant, the
simplest explanation of accelerated expansion, has a checkered
history \cite{Weinberg:1988cp}, \cite{Carroll:1991mt}, having been invoked and subsequently withdrawn several times before. In quantum field theory, we estimate the value of the cosmological 
constant as the zero-point energy with a short-cut scale, for example the Planck scale, 
which results in an excessively greater value than the observational results.

Although the $\Lambda$CDM model has been confirmed as the one that
best fits all the observational tests \cite{h0planck}, in recent years the
precision of measurements and the increase in the number of
them have led to a model \textit{extension} being seriously considered
\cite{valentino}. Among the different ways in which we can deform the 
$\Lambda $CDM model are:

(i) To propose models where the cosmological constant is dynamic, i.e. it changes with time. This family includes the models of quintessence, for example, and from which comes the name {\it dark energy}, which is interpreted as a contribution to the matter content of the universe, th nature of which is unknown.

(ii) Models where the gravitational theory is modified, i.e., it is expected to account for the effect of the cosmological constant.

(iii) Models where one of the fundamental principles of
cosmology is relaxed, which is the homogeneity, openly violating the Copernican
principle.

One of the type (i) models that has received much attention in recent 
years is the model of interaction between dark matter (DM) and DE \cite{Amendola:1999er}, \cite{Zimdahl:2001ar}, \cite{Chimento:2003iea}, \cite{delCampo:2006vv}. Since we do 
not know the nature of DM (non-baryonic) and DE, it is not unreasonable to assume 
that the two may be related. An exploratory form in which this occurs is by assuming 
that there is a small transfer of energy among DE and DM, modeled as a small coupling because the concordance model -- a cosmological constant plus DM -- is a good fit to the data. This scenario, a direct interaction between these dark contributions appears as an observational viable option \cite{salvatelli}.

The interaction is usually modeled phenomenologically by modifying the conservation 
equations through a $Q$ function,
\begin{eqnarray}\label{intsys}
\dot{\rho}_c + 3 H \rho_c = Q, \\ \nonumber
\dot{\rho}_{d} + 3 H (1+\omega)\rho_{d} = -Q,
\end{eqnarray}
where $\rho_d$ is the DE density, $\rho_c$ the DM component, in such a way that only 
the sum of the contributions is conserved, but not each one separately. If $ Q <0 $
there is an energy transfer from DM to DE, and the opposite occurs for 
$Q>0$. Although the sign of $Q$ is not defined from first principles, there are arguments in favor of an overall transfer of energy from DE to DM \cite{Wang:2007ak}, \citep{Abdalla:2007rd},\cite{Yang:2018euj} and also in the other way \cite{Boehmer:2008av}, \cite{Kumar:2017dnp} with also some works indicating a redshift dependence in the sign of $Q$ \cite{Pan:2016ngu}. Mention apart is the result of \cite{Pavon:2007gt} where from a pure thermodynamical argument is demonstrated that the transfer should go from DE to DM. See the review \cite{review} for more details and references about interaction models.

This paper is organized as follows: first we present the three interacting models that will be used. In the second part, the data that will constrain these models is explained, as well as the reason why we chose these data and the different combinations in the analysis. In section III we present the analysis and finally, results and discussions.

\section{The models}

In this paper we study the restrictions on interaction models imposed by observational 
data. Here we study three models of interaction. Explicitly, we study the following 
cases: 
\begin{enumerate}[(i)]
\item $ Q_1 = 3\gamma H \rho_{d}$, 
\item $ Q_2 = 3\gamma H \rho_{c}$,  
\item $ Q_3 = 3\alpha H (\rho'_{d} + \rho'_c)$, 
\end{enumerate}
where in the last case a prime $'$ means a derivative with respect to $\ln a^3$, with $a$ being the scale factor. Both (i) and (ii) were already studied in \cite{xia}. Model (iii) was 
studied first in \cite{chimento} for the case $\omega=-1$ and recently in \cite{Sharov:2017iue}. 
If $ \gamma $ (or $\alpha$) is zero, then there is no interaction. Also, if $ \gamma <0 $ 
(or $\alpha<0$), this indicates that there is transfer of energy from DM to DE.

For the first model (i), the Hubble function $H(z)/H_0= E(z) $ is given by
\begin{eqnarray}\label{ezq1}
E^2(z) = \Omega_m (1 + z)^3 + \Omega_r (1 + z)^4 + \\ \nonumber
+ \Omega_{d}\left( \frac{\gamma}{w + \gamma}(1+z)^3 +
\frac{w}{w+\gamma} (1+z)^{3(1+w+\gamma)} \right),
\end{eqnarray}
where $\Omega_r = 2.469 \times 10^{-5} h^{-2}(1 + 0.2271 N_{\rm eff})$ and $N_{\rm eff}
= 3.04 $, and $ \gamma $ is the parameter that makes the interaction
manifest. Here $ \Omega_m = \Omega_c + \Omega_b $, where $\Omega_c$ is the non-baryonic 
part and $\Omega_b$ is the baryonic one.

For the second model (ii), we obtain
\begin{eqnarray}\label{ezq2}
E^2(z) = \Omega_d (1+z)^{3(1+w)} + \Omega_r (1+z)^4 + \Omega_b (1+z)^3 + \\ \nonumber
+ \Omega_{c}\left( \frac{\gamma}{w + \gamma}(1+z)^{3(1+w)} +
\frac{w}{w+\gamma} (1+z)^{3(1-\gamma)} \right).
\end{eqnarray}
Here the free parameters are $h$, $\Omega_b$, $\Omega_c$, $w$ and $\gamma$.
It is clear that for $ \gamma = 0 $ both expressions - those for models (i) and (ii) - 
reduced to that of the wCDM model.

For the third model (iii) assuming a constant $\omega$, we obtain for $\rho=\rho_c +\rho_{d}$ 
the solution
\begin{equation}
\rho (a) = C_1 a^{3 \beta^{+}} + C_2 a^{3 \beta^{-}},
\end{equation}
where 
\begin{equation}\label{beta}
\beta^{\pm} = \frac{-2-(1-\alpha)w \pm \sqrt{(1-\alpha)^2w^2-4\alpha w}}{2},
\end{equation}
then, the Hubble function can be written as
\begin{eqnarray}\label{ezq3}
E^2(z) = \Omega_b (1+z)^3 + \Omega_r (1+z)^4 + \\
+ (1+z)^{-3\beta^{+}}F_{-} - (1+z)^{-3\beta^{-}}F_{+} . \nonumber
\end{eqnarray}
where
\begin{equation}\label{efes}
F_{\pm} = \frac{\Omega_x (1+w+\beta^{\pm}) + \Omega_c (1+\beta^{\pm})}{\beta^{-} -\beta^{+}}.
\end{equation}
As it is easy to check, turning off the interaction $\alpha=0$, we get from (\ref{beta}) 
that $\beta^{+} = -1$ and $\beta^{-} = -(1+w)$. Replacing in (\ref{efes}), we get 
$F_{+}= - \Omega_x $ and $F_{-} = \Omega_c$ and (\ref{ezq3}) reduces to that of the 
wCDM model.

\section{The data}

In this work, we test the models described in the previous section
using 5 types of data: measurements from $ H(z) $, from type Ia
supernova (SNIa), baryonic acoustic oscillations (BAO), gas mass fraction in clusters $f_{gas}$ and from Cosmic Microwave Background Radiation (CMBR).

Measurements of the Hubble function, $ H(z) $ are taken from several works. They
consist of 31 data points compiled in \cite{Magana:2017nfs} in the range $ z = 0.07 $
and $ z = 1.965 $. We have considered points from \cite{zhang2014}, \cite{Stern2010}, \cite{moresco2012} and also from \cite{moresco2015}. We have used only those $H(z)$ measurements obtained using the differential age method \cite{dem}, and we have exclude those obtained using the clustering method, because we are also using data from BAO.

The data from SNIa are from the Pantheon sample \cite{Scolnic:2017caz}, where the function to be
minimized is
\begin{equation}\label{chi2jla}
\chi^2 = (\mu - \mu_{th})^{T} C^{-1}(\mu - \mu_{th}).
\end{equation}
Here $\mu_{th} = 5 \log_{10} \left( d_L(z)/10pc\right) $ gives the distance modulus where $d_L(z)$ is the luminosity distance, $ C $ corresponds to the covariance matrix delivered in
\cite{Scolnic:2017caz}, and the modular distance is assumed to take the shape
\begin{equation}\label{mujla}
\mu = m - M + \alpha X - \gamma Y,
\end{equation}
where $ m $ is the maximum apparent magnitude in band B, $ X $ is
related to the widening of the light curves, and $ Y $ corrects the
color. In general, cosmology (specified by $\mu_{th}$) is restricted along with the parameters $ M $, $ X $ and $ Y $. The authors of \cite{Scolnic:2017caz} also deliver a binned sample where only $ M $ is a free parameter.

In addition, we used data from BAO compiled in \cite{evslin}.
This set consists of a sample that combines BAO observations from the 6dF survey \cite{2011MNRAS.416.3017B} at redshift $z=0.106$, with distance measurements from 
the Sloan Digital Sky Survey (SDSS) data release 7 (DR7), BAO \cite{Ross:2014qpa} at 
redshift $z=0.15$, and with data from the Baryon Oscillation Spectroscopic Survey 
(BOSS) at redshifts $z=0.32$, $z=0.57$ and $z=2.34$. From the observations it is
possible to measure the BAO scale in the radial and tangential directions, proving 
measurements of the Hubble parameter $H(z)$ and the angular diameter distance 
$D_A(z)$ simultaneously.

At low redshift it is not possible to disentangle the BAO scale in the transverse 
and line-of-sight directions. The BAO observations give the observed ratios of
\begin{equation}\label{dars}
\frac{D_A(z)}{r_s} = \frac{P}{(1+z)\sqrt{-\Omega_k}}\sin \left(\sqrt{-\Omega_k}\int_0^z \frac{dz}{E(z)}\right), 
\end{equation}
for the transverse direction, where $r_s$ is the co-moving sound horizon which is 
independent of $z$, and according to Planck it takes the value $r_s=1059.68$ \cite{planck},
and the ratio
\begin{equation}\label{dhrs}
\frac{D_H(z)}{r_s} = \frac{P}{E(z)}, 
\end{equation}
for the line-of-sight direction. Both in (\ref{dars}) and (\ref{dhrs}) $P=c/(r_sH_0)$, 
which takes the value $30.0 \pm 0.4$ for the best $\Lambda$CDM Planck fit. This parameter 
was used in \cite{evslin} to perform an unanchored BAO analysis, which does not use a 
value for $r_s$ obtained from a cosmological constant, also performed in \cite{aubourg}. 

At low redshift, the surveys give the value for the ratio $D_V(z)/r_s$, where
\begin{equation}
D_V(z) = \left[z(1+z)^2D_A(z)^2D_H(z)\right]^{1/3},
\end{equation}
which is an angle-weighted average of $D_A$ and $D_H$. From \cite{evslin} the data 
considered are: at low redshift, at $z=0.106$ we have $D_V/r_s=2.98 \pm 0.13$, and 
for $z=0.15$, $D_V/r_s = 4.47 \pm 0.17$. For high redshift we consider 
$0.00874D_H/r_s + 0.146D_A/r_s = 1.201 \pm 0.021$ and $0.0388D_H/r_s-0.0330D_H/r_s = 0.781 \pm 0.053$ at $z=0.32$; $0.0158D_H/r_s+0.101D_A/r_s = 1.276 \pm 0.011$ and $0.0433D_H/r_s-0.0368D_H/r_s = 0.546 \pm 0.026$ at $z=0.57$. Following \cite{evslin}, in order to use the BAO 
measurements for the Lyman $\alpha$, we used the $\chi^2$ files supplied on 
the website \cite{baofit} directly. In what follows, we take the Planck value for $r_s$ and 
use $P$ as a function of $H_0$.

We have also used data from gas mass fraction in clusters, $f_{gas}$ as suggested by \cite{Sasaki:1996zz}. In particular we use the data from \cite{Allen:2007ue} which consist in 42 measurements of the X-ray gas mass fraction $f_{gas}$ in relaxed galaxy clusters in the redshift range $0.05<z<1.1$. The $f_{gas}$ data are quoted for a flat $\Lambda$CDM reference cosmology with $h=H_0/100$ km s$^{-1}$Mpc$^{-1}=0.7$ and $\Omega_m=0.3$. To determine constraints on cosmological parameters we use the model function \cite{Allen:2004cd}
\begin{equation}\label{fgas}
f_{gas}^{\Lambda CDM}(z)=\frac{b \Omega_b}{(1+0.19\sqrt{h})
\Omega_M} \left[\frac{d_A^{\Lambda CDM}(z)}{d_A(z)} \right]^{3/2},
\end{equation}
where $d_A(z)$ is the angular diameter distance, $b$ is a bias factor which accounts that the baryon fraction is slightly lower than for the universe as a whole. From \cite{eke98} it is obtained $b=0.824 \pm 0.0033$. In the analysis we also use standard priors on $\Omega_b h^2 = 0.02226 \pm 0.0023$ and $h=0.678 \pm 0.009$ \cite{PDG}. It is important to notice that although the data has been produced using the $\Lambda$CDM as a reference model, to use this data against other models we have to rebuild the data by dividing by the factor $d_A^{\Lambda CDM}(z)$ of the equation (\ref{fgas}).

Finally we use constraints from measurements of the CMB from the acoustic scale $l_A$, the shift parameter $R$, and the decoupling redshift $z_{*}$. The $\chi^2$ for the CMB data is constructed as
\begin{equation}\label{cmbchi}
 \chi^2_{CMB} = X^TC_{CMB}^{-1}X,
\end{equation}
where
\begin{equation}
 X =\left(
 \begin{array}{c}
 l_A - 302.40 \\
 R - 1.7246 \\
 z_* - 1090.88
\end{array}\right).
\end{equation}
The acoustic scale is defined as
\begin{equation}
l_A = \frac{\pi r(z_*)}{r_s(z_*)},
\end{equation}
and the redshift of decoupling $z_*$ is given by \citep{husugi},
\begin{equation}
z_* = 1048[1+0.00124(\Omega_b h^2)^{-0.738}]
[1+g_1(\Omega_{m}h^2)^{g_2}],
\end{equation}
\begin{eqnarray}
g_1 & = & \frac{0.0783(\Omega_b h^2)^{-0.238}}{1+39.5(\Omega_b
h^2)^{0.763}}, \\
 g_2 & = & \frac{0.560}{1+21.1(\Omega_b h^2)^{1.81}},
\end{eqnarray}
The shift parameter $R$ is defined as in \citep{BET97}
\begin{equation}
R = \frac{\sqrt{\Omega_{m}}}{c(1+z_*)} D_L(z).
\end{equation}
$C_{CMB}^{-1}$ in Eq. (\ref{cmbchi}) is the inverse covariance
matrix,
\begin{equation}
C_{CMB}^{-1} = \left(
\begin{array}{ccc}
3.182 & 18.253 & -1.429\\
18.253 & 11887.879 & -193.808\\
-1.429 & -193.808 & 4.556
\end{array}\right).
\end{equation}
Although these priors are obtained using the $\Lambda$CDM as a reference model, they can be used to test models not too far from $\Lambda$CDM. In fact, as we mentioned in the introduction, in this paper we are studying departures from the concordance model assuming a small coupling between DE and DM, so we expect this constrains being useful to put under stress these interacting models. For more details of the work with the data see \cite{Cardenas:2014}.

\section{Results}

As we mentioned before, at low redshift it is not possible to disentangle the BAO scale in the transverse and line-of-sight directions, and therefore the surveys report only the average $ D_V $, usually calibrated using CMB data. At the same time, these low redshift measurements have been consistently in agreement with the $\Lambda$CDM model. However, high redshift BAO detection seems to be at variance with the $\Lambda$CDM model from nearly $2.5 \sigma$ to $3\sigma$. Here, we want to study these effects on three models that present interaction between DE and DM, using not only that for intermediate redshift as $z=0.57$, but also the high redshift ones at $z=2.34$ and $z=2.36$. In the latter cases we made use of the entire likelihood provided by the collaboration \cite{baofit} avoiding the Gaussian approximation used in the literature.

For the analysis we use an Affine-invariant Markov chain Monte Carlo (MCMC) method provided in the {\bf emcee} Python module \cite{emcee} for the five data sets we have mentioned in the previous section. We consider a burn-in phase where we monitoring the autocorrelation time ($\tau$) and set a target number of independent samples. Then, we set 10000 MCMC steps (N) with a number of walkers in the range between 50 and 100. Our estimations of the autocorrelation times for each parameter in the three models all satisfies the relation $N/\tau \gg 50$ suggested in \cite{emcee}, a condition that is consider a good measure of assets convergence in our samplings.
\subsection{Model (i)}

For this model the free parameters are $\Omega_c$, $\Omega_b$, $\omega$, $h$ and $\gamma$.  Although we have performed the analysis for several combinations of data, we presented here only three combinations: A using SNIa+Hz+BAO data, B using SNIa+$H(z)$+BAO+$f_{gas}$, and C using SNIa+$H(z)$+BAO+$f_{gas}+$CMB. The best fit of these parameters from our analysis are display in Table \ref{tab:table01}.
\begin{table}[h!]
\begin{center}
\begin{tabular}{cccc}
\hline
\\
 & $A$ &  $B$ & $C$  \\
\hline
$h$            &  $0.69 \pm 0.02 $   &  $0.687 \pm 0.008$  &  $0.671\substack{+0.01 \\ - 0.008}$    \\
$\Omega_c$   &   $0.25\substack{+0.032 \\ -0.028}$  &   $ 0.293 \pm 0.006 $   &  $ 0.300 \pm 0.004 $  \\
$\Omega_b$   &   $ 0.042\substack{+0.006 \\ -0.005} $ &    $ 0.047 \pm 0.001 $   &  $ 0.048 \pm 0.001$   \\
$\omega$   &  $-1.01\substack{+0.04 \\ - 0.06}$   &   $-1.08 \pm 0.02$   &  $-1.05 \pm 0.02$    \\
$\gamma$   &   $0.03\substack{+0.05 \\ -0.06}$   &  $0.09 \pm 0.01$ &  $ 0.07 \pm 0.005$ \\

\hline
\end{tabular}
\end{center}
\caption{Best fit values of the cosmological parameters for the interaction model (i) using different data sets. A using SNIa+$H(z)$+BAO data. B using SNIa+$H(z)$+BAO+$f_{gas}$, and C using SNIa+$H(z)$+BAO+$f_{gas}+$ CMB. \label{tab:table01}}
\end{table}
Using only SNIa data, or combinations like SNIa+$H(z)$, or SNIa+$f_{gas}$ all indicate a preference for zero interaction, $\gamma \simeq 0$. In Figure 1, the 1$\sigma$, 2$\sigma$ and 3$\sigma$ confidence boundaries for the free parameters of the model (i) are shown, using all the data.
\begin{figure}
\includegraphics[width=15cm]{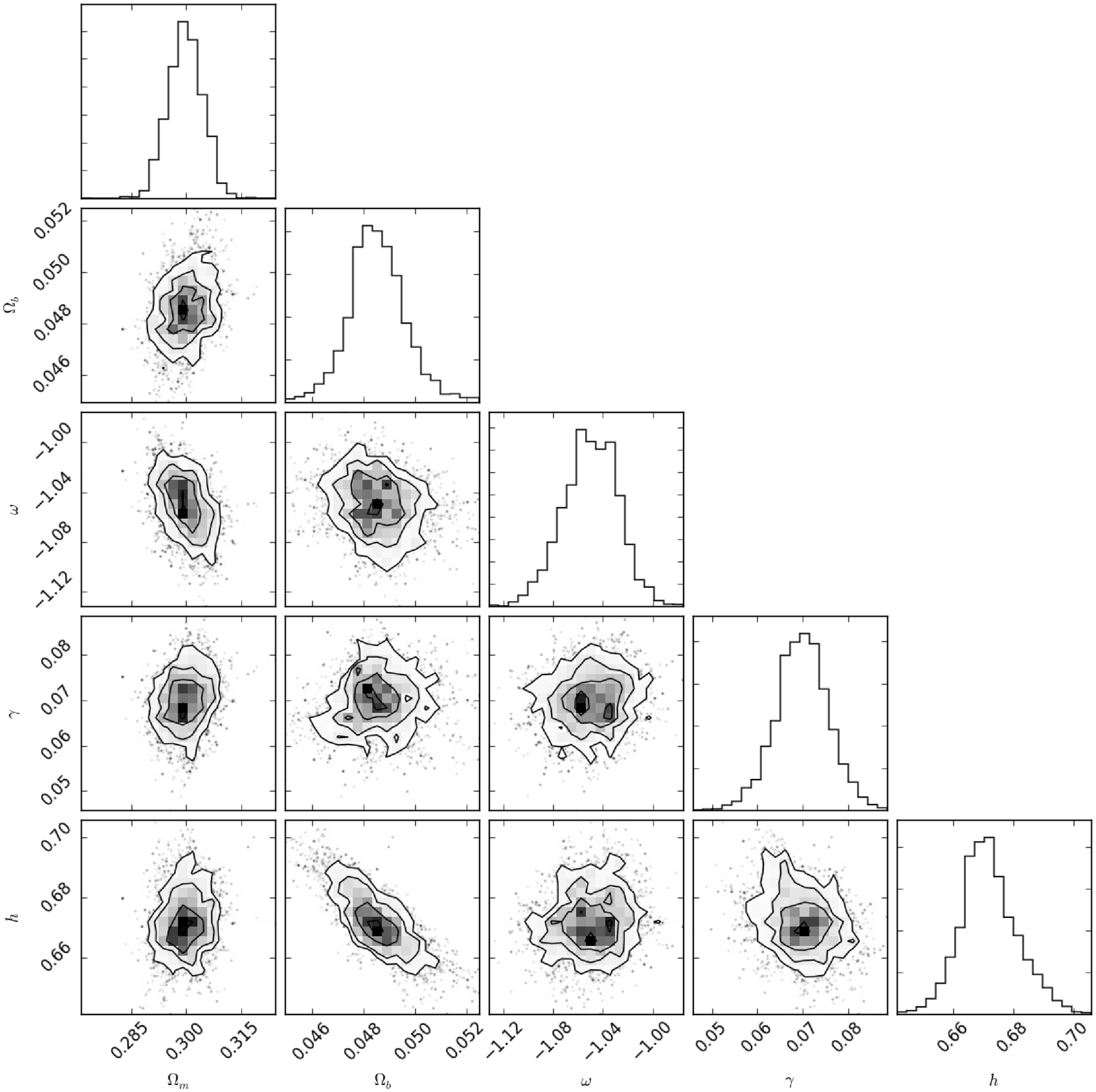}
\caption{We display the results for $ 1 \sigma $, $ 2
\sigma $ and $3 \sigma$ for the model (i) in the parameter space $( \Omega_m, \Omega_b, \omega, \gamma, h )$ using all the data.} 
\end{figure}\label{fig: fig1}
According to the Figure, the constraints imposed by the data are consistent with $ \gamma > 0 $ -- both in the case B and C -- indicating positive evidence for an interacting model.

\subsection{Model (ii)}

\begin{figure}[h]
\centering
\includegraphics[width=15cm]{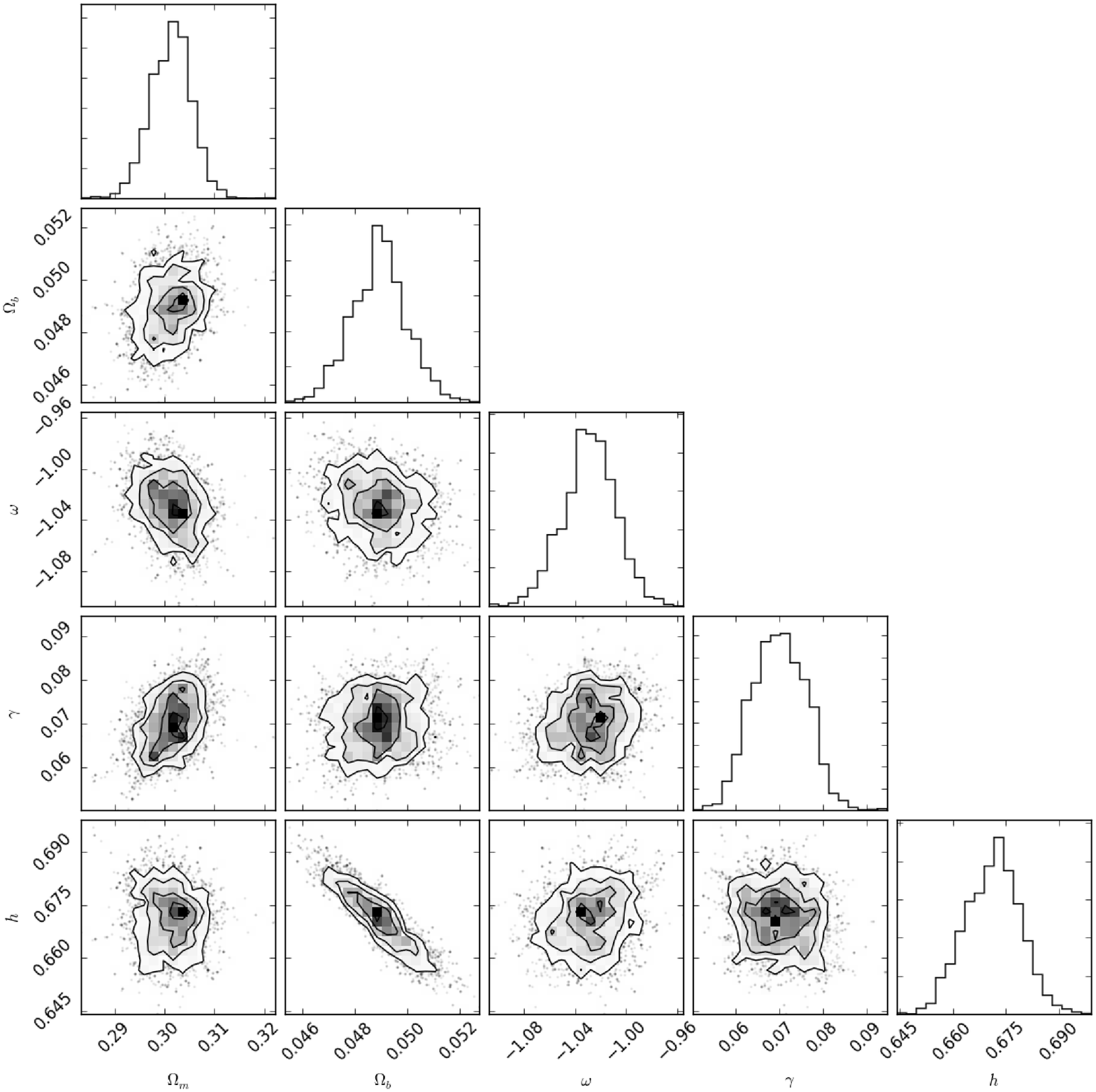}
\caption{We display confidence boundaries for $ 1 \sigma $, $ 2
\sigma $ and $ 3 \sigma $ for the model (ii) for the free parameters $( \Omega_m, \Omega_b, \omega, \gamma, h )$ using all the data.} \label{fig: fig2}
\end{figure}

Next, we show the results of our analysis for model (ii). As in the previous case we presented only the results for the same three configurations mentioned. The results are summarized in Table II and also we plot the best fit contours for the case C, with all the data.
As we can see, the best fit for this case looks quite similar to the previous one. First of all, there is no clear evidence for interaction in the case of the A set of data, but a positive evidence for interaction -- with $\gamma >0$ -- for the cases B and C. Figure \ref{fig: fig2} shows the results for the case C.
\begin{table}[h!]
\begin{center}
\begin{tabular}{cccc}
\hline
\\
 & $A$ &  $B$ & $C$  \\

\hline
$h$ &   $0.682\substack{+0.007 \\ -0.009}$  & $0.682\substack{+0.007 \\ -0.008}$  & $0.669 \pm 0.008$    \\
$\Omega_c$   &   $ 0.29 \pm 0.02 $  &    $ 0.296\substack{+0.005 \\ -0.006}$  & $ 0.301 \pm 0.004 $  \\
$\Omega_b$   &   $0.048 \pm 0.001$   &   $0.048 \pm 0.001$    &  $0.049 \pm 0.001$   \\
$\omega$   &  $-1.01\substack{+0.03 \\ -0.04}$   &   $-1.02\substack{+0.04 \\ -0.03}$   &  $-1.03 \pm 0.02 $    \\
$\gamma$   &   $ 0.004 \pm 0.002$   &  $ 0.003\substack{+0.003 \\ -0.002}$ &  $ 0.071 \pm 0.006 $ \\

\hline
\end{tabular}
\end{center}
\caption{Best fit values of the cosmological parameters for the interaction model (ii) using different data sets. The meaning of A, B and C is the same as in Table I.  \label{tab:table02}}
\end{table}

Comparing between Tables I and II, we note that the best fit values of the parameters are essentially the same. In both cases, as we add more data sets, the best fit for $h$ decreases, $\Omega_c$ increase, $\Omega_b$ remains almost fixed, which is expected because of the prior used, the best fit for $\omega$ is almost the same for each case. Finally for the interaction parameter $\gamma$, in the model (i) the best fit remains almost constant as we add more data sets in the analysis, but the uncertainty diminish. In the case of model (ii) the best fit for $\gamma$ is close to zero for case A and marginally different from zero for case B, but it is definitely not zero for case C.
%
We again observe a non-zero $ \gamma $ using all the data, with a confidence beyond $3 \sigma$. The rest of the parameters fits values rather similar to that of $\Lambda $CDM (Table II). Then, the use of all the data seems to indicate evidence for interaction using model (ii).

\subsection{Model (iii)}

\begin{table}[h!]
\begin{center}
\begin{tabular}{cccc}
\hline
\\
 & $D$ &  $E$ & $F$  \\
\hline
$h$          &   $0.69 \pm 0.02 $  &  $0.678 \pm 0.008$    & $0.680 \pm 0.009$    \\
$\Omega_c$   &   $ 0.36 \pm 0.06 $   &  $ 0.245 \pm 0.005 $   & $ 0.245 \pm 0.006$  \\
$\Omega_b$   &   $0.046 \pm 0.003$   &   $0.048 \pm 0.001$   &  $0.048 \pm 0.001$   \\
$\omega$     &  $-1.4 \pm 0.2$      &   $-1.12 \pm 0.06$    &  $ -1.09 \pm 0.04 $    \\
$\alpha$     &   $ -0.14 \pm 0.12$      &  $ 0.15 \pm 0.06$     &  $ 0.11 \pm 0.03$ \\
\hline
\end{tabular}
\end{center}
\caption{Best fit values of the cosmological parameters for model (iii) using different data sets. The meaning of $D$ is for the combination SNIa+$H(z)$, for $E$ the combination SNIa+$f_{gas}$ and for $F$ the combination SNIa+$H(z)$ + $f_{gas}$ } \label{tab:table03}
\end{table}

Let us study now the third interacting model. As it is clear from the definition (\ref{intsys}), this model is of a different type as those previously studied. The fact that $Q$ depends on the derivatives of the energy densities, makes it naturally a model more difficult to constraint. This observation is also supported by the form of the solution (\ref{ezq3}) compared to the previous ones. As we mention for model (i), we have also performed the analysis using only SNIa data, or combinations like SNIa+$H(z)$, SNIa+$f_{gas}$ or SNI+$H(z)$+BAO. Because it is interesting to discuss these results, we display the best fit values in Table III.
\begin{table}[h!]
\begin{center}
\begin{tabular}{cccc}
\hline
\\
 & $A$ &  $B$ & $C$  \\
\hline
$h$          &   $0.70 \pm 0.02 $  &  $0.70 \pm 0.02$    & $0.69 \pm 0.02$    \\
$\Omega_c$   &   $ 0.37\substack{+0.05 \\ -0.06} $   &  $ 0.37 \pm 0.06 $   & $ 0.31\substack{+0.06 \\ -0.05}$  \\
$\Omega_b$   &   $0.045\substack{+0.03 \\ -0.02}$   &   $0.045 \pm 0.002$   &  $0.046\substack{+0.002 \\ -0.003}$   \\
$\omega$     &  $-1.4 \pm 0.2$      &   $-1.4\substack{+0.2 \\ -0.3}$    &  $ -1.2 \pm 0.2 $    \\
$\alpha$     &   $ -0.1 \pm 0.1$      &  $ -0.15\substack{+0.13 \\ -0.14}$     &  $ 0.0 \pm 0.1$ \\
\hline
\end{tabular}
\end{center}
\caption{Best fit values of the cosmological parameters for model (iii) using different data sets. The meaning of A, B and C is the same as in Table 1.  \label{tab:table04}}
\end{table}
\

As can be see there, in the case D where the data used is the combination SNIa+$H(z)$, the interaction parameter $\alpha$ seems to be centered at a negative value ($\gamma <0$), at least at 1$\sigma$. The rest of the parameters are not too far from those of $\Lambda$CDM except $\Omega_c$ that has clearly a higher value. For the case E --  using the combination SNIa+$f_{gas}$ -- the interaction parameter $\alpha$ shows a clear tendency for a positive value, and different from zero even at 3$\sigma$. The rest of the parameters are those typically of the $\Lambda$CDM. Finally, for the case F -- using the combination SNIa+$H(z)$ + $f_{gas}$ -- the results are almost similar to those of case E, finding evidence at 3$\sigma$ for a positive value of $\alpha$ with the rest of the parameters similar to the $\Lambda$CDM model.

\begin{figure}[b]
\centering
\includegraphics[width=15cm]{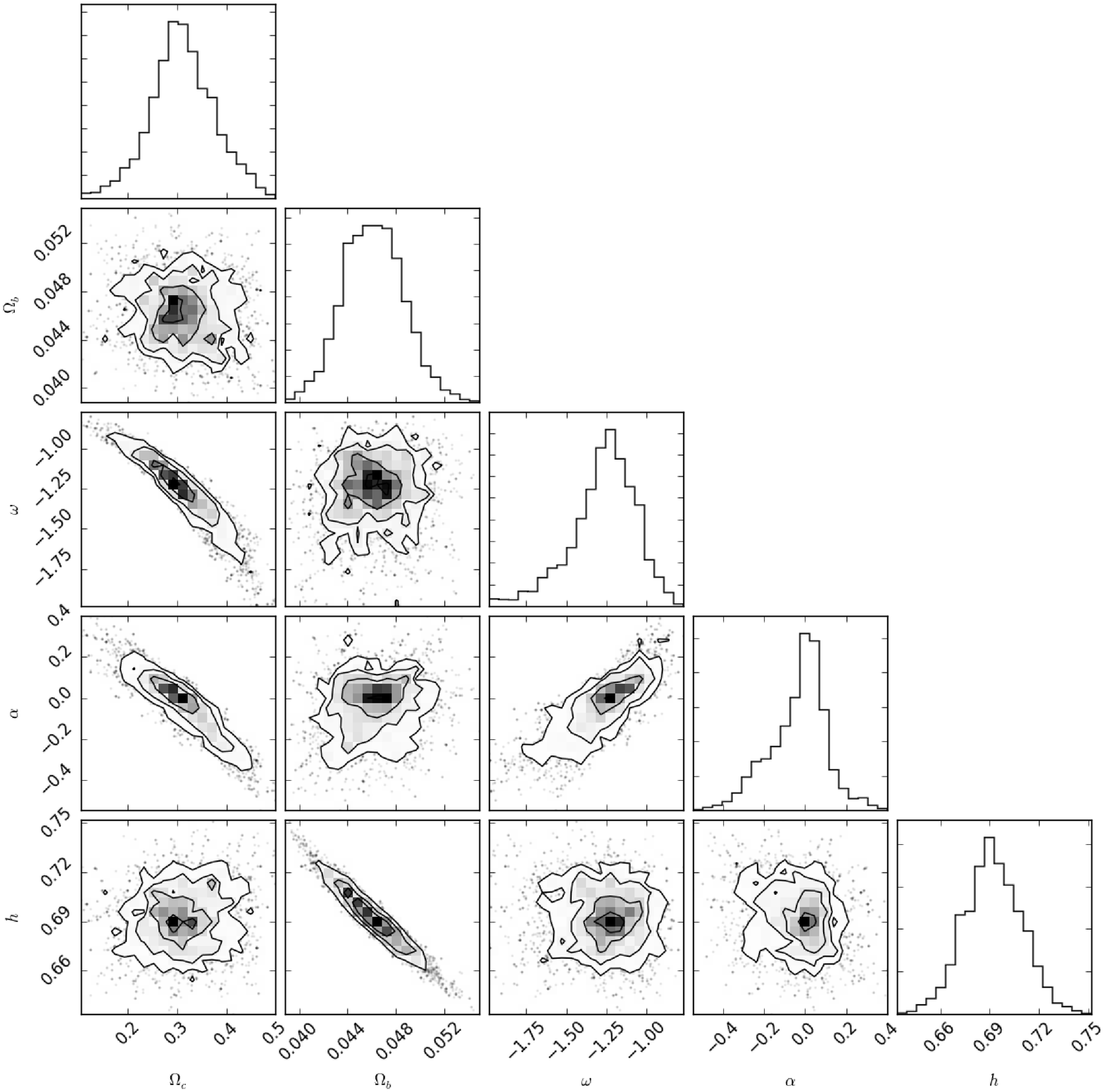}
\caption{We display confidence boundaries for $ 1 \sigma $, $ 2
\sigma $ and $ 3 \sigma $ for model (iii) for the free parameters $( \Omega_c, \Omega_b, \omega, \alpha, h )$ using all the data.} \label{fig: fig3}
\end{figure}

Adding the BAO data to the previous case, the best fit results change a lot. First of all, the best fit of the interaction parameter $\alpha$ moves to a negative value, at least to 1$\sigma$. The value for $\Omega_c$ increases its value but increase also its uncertainty. The $\omega$ parameter moves to a more negative value at 3$\sigma$ from the $\Lambda$CDM value $-1$. All these results are shown in the second column (case B) of Table IV. The first column, the case A for the combination SNIa+$H(z)$+BAO shows almost the same values as that of case B. However, the case C, the one with all the data shows clearly no evidence for interaction, while the rest of the parameters are typically those from the $\Lambda$CDM model.

%
%
\section{Discussion}

We reported the results of analyzing three interaction models between DE and DM, using five observational probes: type Ia supernova, $H(z)$ measurements, BAO data, gas mass fraction in clusters data $f_{gas}$ and CMBR data. In particular we use the BOSS BAO measurements at $z \simeq 0.32$, $0.57$ and $2.34$, using the full 2-dimensional constraints on the angular and line of sight BAO scale.
Using the combination of all the data, the models (i) and (ii) show positive evidence for the existence of an interaction already at 1$\sigma$ -- see Figures (1), and (\ref{fig: fig2})) -- and extended up to 3$\sigma$, implying a transfer of energy from DE to DM, as thermodynamics considerations seems to indicate \cite{Pavon:2007gt}, \cite{review}. However, as we mentioned in the last section for model (iii), using all the data there is no evidence for interaction (see Fig.(3)). 

Although we have used prior for $h$ and $\Omega_bh^2$ and with it, the best fit parameters tend to take values similar to those of the $\Lambda$CDM, in the case of models (i) and (ii) the evidence is clear for a preference of an interacting model with $\gamma >0$ that translate in $Q>0$ indicating a transfer of energy from DE to DM.

Although this study does not incorporate dynamical constraints, such as those from perturbations, our results seem to indicate a chance to find evidence for a non-zero interaction term in the recent cosmological evolution. A work on this topic is under development.


\end{document}